\documentclass[aoas,MSNbibl,nameyear,dvips]{arximspdf}
\usepackage{graphicx}
\doi{10.1214/12-AOAS619} 
\volume{7}
\issue{2}
\pubyear{2013}
\firstpage{1040}
\lastpage{1073}

\makeatletter
\newcommand{\eqref}[1]{(\ref{#1})}
\renewcommand{\citep}[1]{(\citeauthor{#1} \citeyear{#1})}
\def\eqdef{\stackrel{\mathrm{def}}{=}}
\newcommand{\hyp}[1]{(H_{#1})}
\newcommand{\hypalt}[1]{(H_1^{#1})}

\def\ind{{\mathbf{1}}} 

\newcommand{\ie}{i.e.}
\def\Rset{{\mathbb{R}}} 
\def\Sset{{\mathbb{S}}} 
\def\Lset{L}
\def
\prob{{\mathbb{P}}} 
\def\esp{{\mathbb{E}}} 
\def\goto{{\rightarrow}}

\def\cK{\mathcal{K}}

\def\plugintxt{\textsc{PlugIn}}
\def\multipletxt{\textsc{Multiple}}
\def\plugin{\plugintxt}
\def\multiple{\multipletxt}
\def\efron{\textsc{NN}}
\def\twopc{\textsc{TwoPC}}

\def\dd{\,\mathrm{d}}
\def\cB{\mathcal{B}}

\def\jk{_{jk}}

\def\testone{$\mathcal{T}_1$}
\def\testtwo{$\mathcal{T}_2$}

\newproclaim{rem*}{Remark}
\newproclaim{ex}{Example}

\makeatother

\begin{document}
\begin{frontmatter}

\title{Testing the isotropy of high energy cosmic rays using spherical
needlets}
\runtitle{Testing the isotropy of high energy cosmic rays}

\begin{aug}
\author[A]{\fnms{Gilles}~\snm{Fa\"y}\corref{}\ead[label=e1]{gilles.fay@ecp.fr}},
\author[B]{\fnms{Jacques}~\snm{Delabrouille}\ead[label=e2]{delabrouille@apc.univ-paris7.fr}},
\author[C]{\fnms{G\'erard}~\snm{Kerkyacharian}\ead[label=e3]{kerk@math.jussieu.fr}}
\and
\author[D]{\fnms{Dominique}~\snm{Picard}\ead[label=e4]{picard@math.jussieu.fr}}
\runauthor{Fa\"y, Delabrouille, Kerkyacharian and Picard}
\affiliation{Ecole Centrale Paris, CNRS and Universit\'e Paris Diderot,
Universit\'e Paris Diderot and Universit\'e Paris Diderot}

\address[A]{G. Fa\"y\\
Laboratoire MAS\\
Ecole Centrale Paris\\
Grande Voie des Vignes\\
92 295 Ch\^{a}tenay-Malabry\\
France\\
\printead{e1}}

\address[B]{J. Delabrouille\\
Laboratoire APC, CNRS UMR7164\\
10 rue A. Domon et L. Duquet\\
75013 Paris\\
France\\
\printead{e2}}

\address[C]{G. Kerkyacharian\\
Universit\'e Paris X-Nanterre\\
CNRS LPMA\\
175 rue du Chevaleret\\
75013 Paris\\
France\\
\printead{e3}}

\address[D]{D. Picard\\
Universit{\'e} Paris Diderot\\
CNRS LPMA\\
175 rue du Chevaleret\\
75013 Paris\\
France\\
\printead{e4}\hspace*{29pt}}
\end{aug}

\received{\smonth{7} \syear{2011}}
\revised{\smonth{12} \syear{2012}}

%
\begin{abstract}
For many decades, ultrahigh energy charged particles of unknown origin
that can be observed from the ground have been a puzzle for particle
physicists and astrophysicists.
As an attempt to discriminate among several possible production scenarios,
astrophysicists try to test the statistical isotropy of the directions of
arrival of these cosmic rays. At the highest energies, they are
supposed to
point toward their sources with good accuracy. However, the
observations are so
rare that testing the distribution of such samples of directional data
on the sphere is nontrivial.
%
%
In this paper, we choose a nonparametric framework that makes weak
hypotheses on the alternative distributions and allows in turn to
detect various and possibly unexpected forms of anisotropy.
We explore two particular procedures. Both are derived from fitting the
empirical distribution with wavelet expansions of densities. We use the
wavelet frame introduced by [\textit{SIAM J. Math. Anal.} \textbf{38} (2006b) 574--594 (electronic)], the
so-called needlets. The expansions are truncated at scale indices no
larger than some ${J^{\star}}$, and the $L^p$ distances
between those
estimates and the null density are computed. One family of tests
(called \multiple) is based on the idea of testing the distance from
the null for each choice of $J=1,\ldots,{J^{\star}}$,
whereas the so-called
\plugin{} approach is based on the single full ${J^{\star
}}$ expansion,
but with thresholded wavelet coefficients.
We describe the practical
implementation of these two procedures and compare them to other
methods in
the literature. As alternatives to isotropy, we consider both very simple
toy models and more realistic nonisotropic models based on Physics-inspired
simulations. The Monte Carlo study shows good performance of the
\multiple{} test, even at moderate sample size, for a wide sample of
alternative
hypotheses and for different choices of the parameter ${J^{\star}}$.
On the 69 most energetic events published by the Pierre Auger
Collaboration, the needlet-based procedures suggest statistical
evidence for
anisotropy. Using several values for the parameters of the methods, our
procedures yield $p$-values below 1\%, but with uncontrolled
multiplicity issues.
The flexibility of this method and the possibility to modify it
to take into account a large variety of extensions of the problem make
it an
interesting option for future investigation of the origin of ultrahigh
energy cosmic rays.
\end{abstract}

%

\end{frontmatter}

\section{Introduction}
\label{sec:introduction}

\subsection{Motivation}
\label{sec:motivation}

It is a common problem in astrophysics to analyse data sets containing
measurements of a number of objects (such as galaxies of a particular
type) or of events (such as cosmic rays or gamma ray bursts)
distributed on the celestial sphere. Each set of such objects or events
can be represented as a collection of positions $X_i = (\theta_i,\phi
_i), i=1,\ldots,n,$ in $\Sset$ the unit sphere of $\Rset^3$.
In many cases, such objects trace an underlying probability
distribution $f$ on the sphere, which itself depends on the physics
which governs the production of the objects and events. Galaxies, for
instance, form in over-densities of a preexisting smooth field of
distribution of matter in the universe, and the study of the statistics
of their distribution has grown into a field of astrophysics by itself
[\citet{2002sgd..book.....M}].

The case of ultrahigh energy cosmic rays (UHECRs) is of particular
interest, and is the main focus of the present work. UHECRs are
particles of unknown origin which arrive at the Earth from apparently
random directions of the sky. These particles interact with atoms of
the upper atmosphere, generating a huge cascade of billions of
secondary particles. The observation of these secondary particles with
appropriate detectors on ground permits the measurement of the
direction of arrival and of the energy of the original cosmic ray.

The existence of cosmic rays has been known for about a century. Such
particles exist with a very wide range of kinetic energies, from few eV
to more than $10^{20}$~eV.\footnote{1~eV = 1 electron Volt $\simeq1.6
\times10^{-19}$ Joule.} Observed cosmic rays are typically ordinary
charged particles (electrons, protons and nuclei), propagating in empty
space, and deflected by galactic magnetic fields. The rate of observed
cosmic rays in the vicinity of the Earth, however, decreases rapidly
with energy. At low energy, the observed cosmic rays are numerous and
their composition is well known. There also exist several known
astrophysical processes responsible for their acceleration, such as
stellar winds for the least energetic ones, to violent phenomena such
as supernovae shock waves at higher energy. At the highest energies ($E
\geq10^{20}$~eV), however, the observed flux is of the order of 1
event per square kilometre per century, which limits the statistics of
observed events to few tens of events (in two decades of observations).
In addition, no understood astrophysical process, involving known
objects, can accelerate particles to such tremendous energies.

Recent observations of ultrahigh energy cosmic rays suggest that they
are ordinary particles, such as protons and nuclei, accelerated in extremely
violent astrophysical phenomena [see \citet{2011ARAA49119K}, for a
recent review on the astrophysics of UHECRs]. However, many alternate
hypotheses concerning their nature and origin have been proposed over
the years [see,
e.g., \citet{1984ARAA..22..425H},
\citet{2004RPPh...67.1663T},
\citet{2005NuPhS.138..465C}].
UHECRs could originate from active galactic nuclei (AGN), or from
neutron stars surrounded by extremely high magnetic fields, or yet from
many other processes. It is also possible that the type and origin of
ultrahigh energy cosmic rays (at energies above $10^{19}$~eV) depend,
at least to some extent, upon the energy at which they are observed.
Indeed, the most energetic cosmic rays cannot propagate very far (i.e.,
not much more than $\sim100 $ Mpc) without losing most of their energy
by interactions with photons from the Cosmic Microwave Background
[the so-called GZK effect; \citet{1966PhRvL..16..748G},
\citet{1966JETPL...4...78Z}].
The confirmation of the energy cutoff at the high end of the cosmic ray
spectrum is one of the main achievements of the Pierre Auger
Observatory [\citeauthor{2008PhRvL.101f1101A} (\citeyear{2008PhRvL.101f1101A,2010PhLB..685..239A})].

Before the location and physical process of acceleration have been
clearly identified, taking into account the fact that most of the
evidence about the chemical composition of cosmic rays at the highest
energies rely on extrapolations of the present knowledge of hadronic
interactions at energies two orders of magnitude above the range
presently tested at the LHC, it is difficult to completely rule out
alternate theoretical explanations as to what UHECRs exactly are and
what is their origin. Alternate hypotheses such as production by decay
of long-lived relic particles from the Big Bang, about 13 billion years
old [\citet{2000PhR...327..109B}], are just starting to be disfavored by
the observations of the Pierre Auger collaboration, with recently
published results about primary photon limits that impose stringent
limits on these kinds of models [\citet{2009APh....31..399P}].

In an attempt to better understand the origin of such UHECRs,
physicists study the statistical distribution of their directions of
arrival, looking for two particular signatures. First, the
(statistically significant) arrival of more than one UHECR from the
same direction on the sky would indicate that their production is not
likely to originate from single time events (e.g., catastrophic mergers
of two compact astrophysical objects), but rather from sources which
emit UHECRs regularly.\footnote{With the caveat that the time of
propagation may depend on the energy and on the exact trajectory
followed by the UHECR to reach us, making it possible that two
particles reaching the Earth at different times have actually been
emitted simultaneously.} Second, one may look for correlation in the
directions of arrival of UHECRs with known astrophysical objects, as
nearby active galactic nuclei, in an attempt to identify plausible
production sites. Hence, in some hypotheses, the underlying probability
distribution for the directions of incidences of observed UHECRs would
be a finite sum of point-like sources---or nearly point-like, taking
into account the deflection of the cosmic rays by magnetic fields. In
other hypotheses, the distribution could be uniform, or smooth and
correlated with the local distribution of matter in the universe. The
distribution could also be a superposition of the above. Distinguishing
between these hypotheses is of primordial importance for understanding
the origin and mechanism of production of UHECRs.

In the past 20 years, a number of experiments have gathered
observations of UHECRs, and several papers have been written which look
for such features in the distribution of their directions of arrival,
with sometimes contradictory conclusions.
The difficulty lies in the fact that UHECRs are rare and that they do
not arrive necessarily exactly from the direction where their source is
located. Indeed, as typical cosmic ray particles are charged (which
permits their acceleration by electromagnetic processes), they are
deflected by Galactic and intergalactic magnetic fields. The deflection
depends on the length of the path through the magnetic field and on the
energy and charge of the particle. In fact, only very energetic cosmic
rays (above few $10^{19}$~eV) with small charge (e.g., protons or
nuclei with small atomic numbers) are expected to travel typical
astrophysical distances from their source to us with deflection angles
smaller than a few degrees. Details of the deflections are not known,
as neither the exact magnitude, orientation and regularity on large
scales of Galactic and extragalactic magnetic fields, nor the distance
of the sources of UHECRs, nor the exact energy of the incoming cosmic
ray, nor its charge (to within a factor of 26 between protons and iron
nuclei), are known. Errors on the direction of the source of an UHECR
can then be of order $1^\circ$ at the lowest (typical error on the
measurement of the direction of arrival with Auger), up to few degrees
for protons, or tens of degrees for heavy nuclei travelling a long path
through a regular galactic magnetic field.

Given a set of observed UHECRs, how can one best test for ``repeaters''
(cosmic rays coming from the same source) or, more generally,
anisotropy in the distribution? If one restricts the analysis to the
few events for which one is sure that the deflection angle is
negligible, events are scarce and there are not enough statistics to
conclude. As one selects events with less energy, the direction of
origin becomes less reliable, with the total number of events
completely dominated by those events with poorly constrained direction
of origin. Finally, it is not clear how to build the isotropy test,
without any sound prior knowledge about the uncertainty in the measured
direction of the source. All of these are very meaningful questions to
analyze UHECR observations.

Recently, an analysis of the direction of arrival of 27 UHECRs observed
by the Pierre Auger experiment concludes in the existence of an
anisotropy and a correlation with objects in a catalogue of nearby
active galactic nuclei (AGNs), located at distances lower than about 70
Mpc\footnote{70 million parsecs $\simeq2.15 \times10^{21}$ km.}
[\citet{2008PhRvL.101f1101A}]. This anisotropy, however, is less obvious in a
more recent analysis, based on 69 observed events [\citet{2010APh....34..314T}].
Clearly, the statistics are limited, and the development of new methods
for investigating this topic can provide new insights on the origin of
the UHECRs. Methods independent of external data sets such as the
forementioned VCV catalogue (which is not a statistically
well-characterized sample of AGNs but a compilation of published
results) are of particular interest.

\subsection{Outline of this work}
\label{sec:outline-paper}

This work focuses on the important question of the isotropy of the
cosmic rays. Because of the small number of available data, this
question is not answered yet, although data from the Pierre Auger
collaboration seems to hint at a correlation between the directions to
the ultra-high energetic events (above $5.5 \times10^{19}$~eV) and the
directions to active galactic nuclei in the catalogue compiled by V\'
eron and Cetty-V\'eron [see \citeauthor{2008APh....29..188P} (\citeyear{2008APh....29..188P,2010APh....34..314T})].
From a statistical
point of view, we address the question of testing the goodness of fit
of the isotropy assumption to this small sample of directional data.
The framework we choose is purely nonparametric, as we do not want to
favour any particular alternative hypothesis, and as we wish to be able
to discover unexpected forms of anisotropy.

The paper is organized as follows. In Section \ref{toymodel} we present
a simplified model of cosmic ray propagation which will be used in
Monte Carlo simulations to test the method. In Section \ref
{sec:nonp-tests-anis} we present the nonparametric framework.
Then we describe our needlet based anisotropy tests in Section \ref
{sec:anisotropy-tests}. In Section \ref{sec:monte-carlo-exper} we
present a Monte Carlo experiment that compares the power of the
different tests and also the robustness of this power with respect to
the parameters of the methods. We apply our procedures to real data
from the Pierre Auger collaboration in Section~\ref{sec:analysis-auger-data}.
We then conclude and give perspectives for future extensions of the
present work.
An online supplement [\citet{fay:etal:2013b}]
is devoted to a longer description of the type of wavelets we have used
(the needlets) and the practical and numerical implementation of our
methods. More numerical results are available there.

\section{Simulating cosmic ray emission}
\label{toymodel}

In our investigation of tools to analyse the distribution of UHECR
events, we need a way to simulate a distribution of observed events as
a function of an underlying physical model. A complete Monte Carlo
simulation of the physical processes of cosmic ray emission and
propagation in the magnetic fields is beyond the scope of this paper
and too dependent on a number of physical assumptions for which there
is little available knowledge. We decide to perform qualitatively
relevant simulations using a simple, although physically
representative, toy model of cosmic ray emission and propagation.

\subsection{Cosmic ray sources}

In one hypothesis $\hyp{0}$, we will assume that cosmic rays are emitted
from a uniform distribution of many sources, that is, their directions
of arrival are independent of the energy, and uniformly distributed on
the celestial sphere. In the alternate hypothesis $\hyp{1}$, we will
assume that $n$ cosmic rays originate from a small number $n_s$ of
sources, distributed uniformly in a spherical volume $V$ of universe,
of radius $r_{\max} = 70$~Mpc. For $n_s \gg n$, the distribution of
directions of origin will be close to uniform and $\hyp{1}$
indistinguishable from $\hyp{0}$.
For $n \gg n_s$, and $n_s$ small, coincidences in the directions of
arrival of the observed UHECRs will permit to identify easily the
directions of the emitting sources. Our objective is to address the
issue when $n_s$ is comparable to the number of observed events $n$.

\begin{figure}

\includegraphics{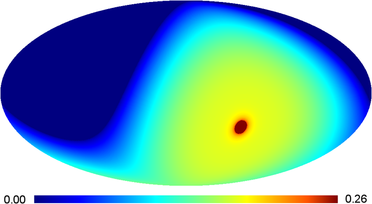}

\caption{Exposure function $\varepsilon$ for the Pierre Auger Observatory
in Galactic coordinates, represented through a Mollweide projection and
computed from geometrical considerations [see \citet{2001APh....14..271S}]. The value of the exposure for some
direction is defined as the probability that an incoming event from
this direction is actually detected by the instrument. See
Section \protect\ref
{sec:NPintroduction}.}
\label{fig:auger_exposure}
\end{figure}

Simulations are performed as follows:

\begin{itemize}
\item We fix the number $n_s$ of sources and distribute them uniformly
in the volume~$V$. We assume that all sources are physically identical,
that is, they emit cosmic rays with the same probability and the same
distribution in energy, the latter coinciding with the observed flux $dN/dE$.
\item We fix the number $n$ of observed cosmic rays and draw at random
their energies according to the distribution $n(E) \propto E^{-\alpha
}$, $E \in[E_{ \mathrm{min}}, E_{ \mathrm{max}}]$, $\alpha> 0$.
\item For each observed cosmic ray, we assign at random a corresponding
emitting source, according to a probability density inversely
proportional to the square of the distance $D$ to the source (sources
nearer produce a larger flux on Earth). This probability distribution
can be modulated by the acceptance of the instrument for studying
realistic test cases. For instance, \citet{2010APh....34..314T} uses 69
highest energy events for the search of correlations with astrophysical
sources, selected by a cut in zenith angle of arrival ($\theta_{\mathrm{zenith}} \leq60^\circ$). Assuming homogeneous time coverage in UT over
the years of observation, the exposure is computed straightforwardly
from simple geometrical considerations [see \citet{2001APh....14..271S} and the details at
the end of Section \ref{sec:defl-galact-extr}].
The map of Auger exposure computed in this way is displayed in Figure
\ref{fig:auger_exposure}. The effect of the GZK cutoff is taken into
account simply by limiting the volume to a sphere of 70 Mpc radius.
\item For each cosmic ray, we modify the direction of arrival due to
extragalactic magnetic fields. The next subsection describes the model
used to implement these deflections.
\end{itemize}

%
%

\subsection{Deflection by Galactic and extragalactic magnetic fields}
\label{sec:defl-galact-extr}

Galactic magnetic fields are an important component of the Galactic
interstellar medium (ISM). They can be probed in a variety of ways.
The impact of local magnetic fields is observed in the optical
wavelength range via starlight polarization. Indeed, elongated interstellar
dust grains in the foreground of the observed star, aligned
perpendicularly to magnetic field lines, absorb preferentially one
direction of starlight polarisation (along their major axis). Measurements of many stars reveal
a general picture of the magnetic field in the Milky Way near the Sun
[\citet{1996ASPC...97..457H},
\citet{2002ApJ...564..762F}]. Aligned dust grains
also emit polarized infrared emission, which can be used to infer
magnetic fields in dust clouds [\citet{2004AA...424..571B}].
Zeeman splitting of radio spectral lines allows for the direct
measurement of relatively strong fields in nearby, dense gas clouds in
the Milky Way [\citet{2010ApJ...725..466C}]. On larger-scales, the
magnetic field of our Galaxy can be probed in three dimensions using
Faraday rotation of pulsar signals [\citet{2006ApJ...642..868H}].
Finally, synchrotron emission, emitted by relativistic electrons
spiralling in the magnetic field, can be used to constrain the
direction and amplitude of the magnetic field either from direct
observation of the synchrotron polarisation
[\citet{2007ApJS..170..335P}]
or by measuring the Faraday rotation of Galactic synchrotron using
multi-wavelength observations in the radio range (below few GHz) [\citet
{2011AIPC.1381..117B}].

In the vicinity of the Sun, the Galactic magnetic field has a typical
amplitude of a few microGauss. This amplitude is typically increasing
with decreasing distance toward the Galactic center, where it can
reach values of a few tens of microGauss, and up to a few milliGauss in
very local regions. In general, the regular component over most of the
outer Galaxy is of the order of a few microGauss, aligned along the
Galactic plane. The overall field structure follows the optical spiral
arms, with evidence for at least one large-scale field reversal in the
disk, inside the solar radius, and several distortions near
star-forming regions.

For the purpose of estimating their impact on the deflection of high
energy cosmic rays, Galactic magnetic fields are typically modeled as
the sum of two components with different physical properties, a regular
component and a turbulent component. The regular component roughly
follows the spiral arms of the Galaxy and induces deflections typically
perpendicular to the Galactic plane, that is, deflections in latitude
of arrival. The turbulent component induces random deflections, which
can be modeled as two-dimensional Gaussian distributions centered at
the source. Indeed, we assume that such deflections are made of the
superposition of many independent small deflections by independent
regions with independent magnetic field directions, so that the
Gaussian hypothesis is justified by the central limit theorem. We
consider only cases in which the total deflection is small enough that
the projection to the sphere is irrelevant (as well as the truncation
of angles to~$2\pi$). Typical deflections for atomic nuclei are as
follows [\citet{2002JHEP...07..006H}].

For the regular component (magnetic lensing effect),
%
\begin{equation}
\label{eq:B_reg} \delta_{\mathrm{reg}} = 3.25^\circ \biggl(
\frac{10^{20}\ {\mathrm{eV}}}{E/Z} \biggr) \biggl( \frac{B}{2\ \mu{\mathrm{G}}} \biggr) \biggl(
\frac{r}{3\
{\mathrm{kpc}}} \biggr),
\end{equation}
where $E$ is the energy of the UHECR in eV, $Z$ is the atomic number
[e.g., 1~for hydrogen nuclei (protons), 2 for Helium nuclei (alpha
articles), etc.], $B$~is the magnetic field in microGauss ($\mu$G), and
$r$ the propagation length of the cosmic ray in the magnetic field. The
deflection is assumed deterministic (although energy-dependent), and
the instantaneous direction of the deflection is along $\vec{v} \times
\vec{B}$, where $\vec{v}$ is the velocity of the incoming particle and
$\vec{B}$ the regular Galactic magnetic field, assumed to be along the
$y$-axis of the Galactic coordinate system.

For the turbulent component (random deflection),
%
\begin{equation}
\delta_{\mathrm{turb}} = 0.56^\circ \biggl( \frac{10^{20}\ {\mathrm{eV}}}{E/Z} \biggr)
\biggl( \frac{B}{4\ \mu{\mathrm{G}}} \biggr) \sqrt{\frac
{r}{3\
{\mathrm{kpc}}}} \sqrt{
\frac{L_{\mathrm{gal}}}{50\ {\mathrm{pc}}}} \label{eq:B_turb}.
\end{equation}
The deflection is Gaussian distributed with a standard deviation
$\delta
_{\mathrm{turb}}$ and uniform distribution of the direction of the deviation
in $[0, 2\pi[$.
The deflections are written in terms of the typical expected values for
the magnetic field (2~$\mu$G for the regular part and 4~$\mu$G for the
turbulent part), for coherence length $L_{\mathrm{gal}}$ of the turbulent
part of the Galactic magnetic field (about 50~pc). 3~kpc is the typical
propagation length $r$ inside the Galactic magnetic field for a cosmic
ray coming perpendicularly to the Galaxy. A plane parallel
approximation of the disc-shaped geometry of the Milky Way suggests a
dependence of $r$ with the Galactic latitude $b$ of the incoming cosmic
ray. We assume here a dependence $r \propto1/\sin b$, with a maximum
length of 10~kpc, typical of the size of the Galactic disk.

Extragalactic magnetic fields also deflect cosmic rays originating from
distant locations in the Universe. These deflections are expected to be
qualitatively similar to those due to the turbulent part of the
Galactic magnetic field, except that typical field strengths are
smaller (and less well known) and correlation lengths are larger.
Following The Pierre Auger collaboration [\citet{2008APh....29..188P}],
we assume a deflection with standard deviation given by
%
\begin{equation}
\delta_{\mathrm{ext}} = 2.4^\circ \biggl( \frac{10^{20}\ {\mathrm{eV}}}{E/Z} \biggr)
\biggl( \frac{B}{1\ {\mathrm{nG}}} \biggr) \sqrt{\frac{D}{100\
{\mathrm{Mpc}}}} \sqrt{
\frac{L_{\mathrm{ext}}}{50\ {\mathrm{pc}}}} \label{eq:B_ext}.
\end{equation}

UHECRs are observed to arrive on Earth with a flux $dN/dE$ proportional
to $E^{-4.2}$ for energies $E > 4 \times10^{19}$~eV
[\citet{2008PhRvL.101f1101A}].
Although the shape of the spectrum is not very well constrained in this
region (more recent Auger results suggest a spectral index closer to
$-4.3$), the exact shape of the spectrum does not have a strong impact
on the validity of our analysis.
Our simulations will assume such a distribution, with various values
for the minimum energy $E_{\mathrm{min}}$ and $E_{\mathrm{max}} = 10^{21}$~eV.
We focus on very energetic UHECRs ($E> 10^{19}$~eV) and assume UHECRs
are light nuclei ($Z\approx1$), for which deflections by magnetic
fields are expected
to be of the order of a few degrees.

We then implement cosmic ray deflections according to equation (\ref
{eq:B_ext}) (first the cosmic ray travels in the intergalactic medium)
and then using both equations (\ref{eq:B_reg}) and (\ref{eq:B_turb}).
As the exact nature of the cosmic rays has little impact on the general
principles of our method, except that a change in atomic number induces
a change in the scale of the deflections, we have assumed here for
simplicity that all cosmic rays are protons (i.e., $Z=1$). This,
however, as a further refinement, can be easily changed for practical
application on real data sets. In particular, the presence or lack of
anisotropy in the directions of arrival of the highest energy cosmic
rays may help shed light on the nature of these particles, as iron
nuclei, for instance, are more deflected by magnetic fields than
protons, by a factor $Z_{\mathrm{iron}} = 26$. This is an important point to
take into account in view of recent Auger results that seem to indicate
a low proton fraction at energy above $10^{18}$~eV, so that the cosmic
rays at those energies might be essentially heavier nuclei [\citet
{2010PhRvL.104i1101A}].

\begin{figure}

\includegraphics{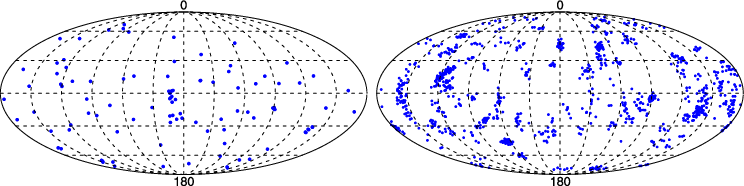}

\caption{Two simulations of the physical model described in
Section \protect\ref
{toymodel}, with $\alpha=4.2$, $E_{\min}=4 \times10^{19}, E_{\max} =
10^{21}$. On the left, the number of sources is $n_s = 1000$ and the
number of observations is $n = 100$. On the right, $n_s = 100$ and $n =
1000$. It appears in this latter case that clusters of events are of
different typical angular size.}
\label{fig:example_distribution}
\end{figure}

Figure \ref{fig:example_distribution} illustrates simulated outcomes in
two extreme cases: few sources and many cosmic rays (right) and many
sources and few cosmic rays (left).

%
%

In practice, instruments observe the sky unevenly. The capability of
the instrument to observe in a particular direction of the sky depends
on the field of view of the instrument and on the orientation of the
instrument with respect to the sky (which itself depends on the
sidereal time). From the properties of the instrument and the geometry
of the observations, one can infer an equivalent observing time as a
function of direction on the sky, that is, a function on the sphere
that modulates the probability of detection of the observed cosmic rays.
As an illustration, we have displayed on Figure~\ref
{fig:auger_exposure} a Mollweide projection of the exposure map
associated with the Pierre Auger Observatory, in Galactic coordinates,
computed following Section 2 in \citet{2001APh....14..271S}. This
exposure map has been generated assuming a maximum accepted zenith
angle for incoming cosmic rays of $\theta_{\mathrm{zenith}}=60^\circ$ and
uniform distribution of observation periods in universal time (and
hence, an exposure that depends exclusively of the declination, not the
right ascension, in equatorial coordinates). The effect of the
precession of equinoxes has been neglected for generating this exposure
map (the perturbations it would generate are very tiny as compared to
what we can measure with about 100 events, as currently available).

\section{Nonparametric tests on the sphere}
\label{sec:nonp-tests-anis}

\subsection{Introduction}
\label{sec:NPintroduction}

We assume that the cosmic rays arrive on Earth along a directional
density $h$. We need to test
{\renewcommand{\theequation}{\testone}
\begin{equation}
\label{eq:T1}
\hyp{0}\dvtx h \equiv h_0\qquad
\mathrm{against}\ \hyp{1} \dvtx h \not\equiv h_0,
\end{equation}}
\hspace*{-2pt}where $h_0$ is the density under the null. For the above-mentioned
reasons, we focus on test for anisotropy, then we choose $h_0 \equiv
\frac{1}{4\pi}$. Note, however, that the whole subsequent setup can
handle anisotropic choices for the null distribution.
To take into account the nonuniform angular acceptance in the
observation model, we model the exposure of the instrument by a known
and arbitrary function $\varepsilon\dvtx\Sset\to[0,1]$. In this setting, we
assume that incoming events from direction $\xi\in\Sset$ have
probability $\varepsilon(\xi)$ to be observed by the instrument.
In this case, the observed incidental directions are distributed along
a density $f$ which is proportional to $\varepsilon h$:
%
\setcounter{equation}{3}
\begin{equation}
f(\xi) = \frac{\varepsilon(\xi) h(\xi)}{\int_\Sset\varepsilon(\xi')
h(\xi')
\dd\xi'} \cdot\label{eq:density_observations}
\end{equation}
Under the null, the observed directions of cosmic rays have a density
\[
g(\xi) = \frac{\varepsilon(\xi) h_0(\xi)}{\int_\Sset\varepsilon(\xi
') h_0(\xi
') \dd\xi'} \cdot
\]
Let $(X_1,\ldots,X_n)$ be an $n$ sample of i.i.d. random positions on the
two-dimensional sphere with probability density function
$f$. In order to test for isotropy of the underlying physical
phenomenon in this observational context, we need to implement the test
{\renewcommand{\theequation}{\testtwo}
\begin{equation}
\label{eq:T2}
\hyp{0}\dvtx f \equiv g \qquad\mathrm{against}\ \hyp{1} \dvtx
f \not\equiv g.
\end{equation}}
On the real line, testing for $f \equiv g$ can be reformulated as
testing for the uniform distribution of the sample $G(X_1),\ldots,G(X_n)$ on $[0,1]$, where $G$ is the distribution function associated
with the probability density $g$. For higher dimensions (as on the
sphere), there is no natural transformation of the data, no notion of
distribution function for directional data, that allows to recast
\eqref{eq:T2} as \eqref{eq:T1}. Then we consider \eqref{eq:T2} in its
generality, with \eqref{eq:T1} as a particular case.

Our aim in this paper is to provide test algorithms which are at the
same time easy to implement, efficient in practical situations where
the sample size is small (a few tens) and the data may be collected in
a nonuniform or incomplete way, but also with properties that are
likely to be optimal from a theoretical point of view.

Let us begin with a short review on nonparametric tests associated to function
estimation, since this will inspire our study in many ways.

\subsection{Anisotropy tests among general nonparametric tests}
\label{sec:anis-tests-among}
The test problem is well posed when the alternative is given. More
often in practice it is wiser to consider a large nonparametric class
of alternatives. To allow derivation of optimality properties,
following standard point of view in a nonparametric framework
[see, e.g., \citet{ingster:suslina:2003},
\citet{ingster:1993abc},
\citet{butucea:tribouley:2006}],
we shall consider smooth alternatives of the form
%
\setcounter{equation}{4}
\begin{equation}
\label{eq:h1n} \hyp{1,n}\dvtx f \in\mathcal F_n(d,C),
\end{equation}
where
%
\begin{equation}
\mathcal F_n(d,C) = \bigl\{ g' \in\mathcal R\dvtx d
\bigl(g'z,g\bigr) > Cr_{n}\bigr\} \label{eq:deffn}
\end{equation}
and $\mathcal R$ is a class of regularity, that contains, for example,
all the twice continuously differentiable densities or densities
satisfying the H\"{o}lder condition with H\"{o}lder exponent $s > 0$.
We may consider balls in Sobolev or Besov spaces (see below). Here, $d$
is a (semi-) distance between densities and $r_{n}$ is referred to as a
\emph{separation rate}. Roughly speaking, $d$ and $r_n$, respectively,
define the shape and the size of the neighbourhood of the density under
the null which is excluded from the alternative set of densities. The
multiplicative constant $C$ allows to define the concept of critical
separation rate; see equations (\ref{eq:rate-optimal-1}) and (\ref
{eq:rate-optimal-2}) below.

The choice of such alternatives is essential for the test procedure
because the test statistics are built, more or less, on estimators of
$d(f,g)$. For some particular distances, nonparametric estimators $\hat
f$ of the density of the observed sample may be plugged into the
distance, namely,
\[
\hat d(f,g) = d(\hat f,g).
\]
For instance, $\hat f$ could be a histogram-like (pixel-wise constant)
density estimate of $f$ based on counting events falling in any pixel
of a given tessellation $\{V_k\}_{k=1,\ldots,K}$ of the sphere, namely,
\[
\hat f = \frac1 n \sum_{k=1}^K \#
\{X_i \in V_k, i=1,\ldots,n\} \frac
{\ind_{V_k}}{\mu(V_k)}
\]
and the decision could be taken on the value of $d(\hat f,g) = \|\hat f
- g\|^2$, say. Nevertheless, as described in \citet{ingster:2000}, such
``plug-in'' procedures are not always optimal in terms of rates of
separation (see Section \ref{sec:plugin-tests} for a more precise
statement). In contrast, multiple tests have nice theoretical (minimax
optimality and adaptivity) properties in various contexts: detection in
a white noise model [\citet{spokoiny:1996}], $\chi^2$ test of uniformity
on $[0,1]$ [\citet{ingster:2000}], goodness-of-fit test and model selection
for random variables on the real line [\citet{fromont:laurent:2006}],
two-sample homogeneity tests [\citet{butucea:tribouley:2006}], for
instance. Note that one would also like to test for uniformity by
taking into account the uncertainty on the measurements of the
directional data. In a first approximation, this error can be modeled
as a convolution noise: the observations are $Z_i = \varepsilon_i X_i,
i=1,\ldots,n,$ where $\varepsilon_1,\ldots,\varepsilon_n$ is an i.i.d. sequence
of random rotations in SO(3). \citet{2012arXiv1203.2008L} addressed the
problem of testing for the isotropy $(X_1,\ldots,X_n)$ in the particular
case of a full-sky coverage with uniform exposure and from noisy
observation (random rotations of the directions). As a consequence of
the uniform coverage, their adaptive testing procedure is ideally
constructed on the multipole moments of the observations.

If one has strong prior information, it is possible to construct tests
that are
not uniform except along a few set of directions, but which can have as much
power as possible at the $n^{-1/2}$ scale in those few directions of
interest. This framework is introduced in \citet{MR2281882} and applied in
\citet{2008ApJ...685..384B} for detecting periodicity in a sequence of photon
arrival times. In our context, those directions are described by the Besov
regularity of the alternative density, which is efficiently handled by the
formalism of the wavelet analysis.

In the following paragraphs we discuss the various ingredients of our study.

\subsubsection{Distances}
\label{sec:distances}
We will consider standard distances of functions on the sphere,
although there is in fact no clear choice for a 'good' distance in this
framework: $\Lset^1$ distance is generally more appropriate for
probability densities, but $\Lset^p$ distances when $p$ is increasing
and especially $\Lset^\infty$ are more and more sensitive to bumps.
As it is both usual and practical, we will mainly consider the $\Lset
^2$ distance (with respect to the invariant measure on the sphere).
But, we will also consider expressing our results for other $\Lset^p$
distances such as $\Lset^1$ and $\Lset^\infty$. It is important to
notice that it is the remarkable ability to concentrate the needlets
that enables us to consider various distances. More traditional bases
would only allow the $\Lset^2$ distance and would then be much less
sensitive to local changes.

\subsubsection{Separation rate}
\label{sec:separation-rate}

Let $T(X_1,\ldots,X_n) \in\{0,1\}$ be a nonrandomized decision, that
is, a measurable function of the sample $(X_1,\ldots,X_n)$ with value in
$\{0,1\}$. The dependence in $n$ is omitted in most of our notation. As
usual the event $[T=1]$ is equivalent to the rejection of the null
hypothesis. The probability of error of the first kind (false positive)
of the decision is denoted
%
\begin{equation}
\label{eq:defleveltest} \alpha_n(T) = \prob_g(T=1),
\end{equation}
while probability of error of the second kind (false negative)
against the alternative~(\ref{eq:h1n}) 
is
%
\begin{equation}
\label{eq:defpowertest} \beta_n(T,C) = \sup_{f\in\mathcal F_n(d,C)}
\prob_{f}(T=0). 
\end{equation}
%
Here $\prob_f, \prob_g$ denote the probability measure under the
density $f$ or $g$ for the i.i.d. sample $(X_1,\ldots,X_n)$.

Formally, the separation rate is defined using the following minimax
optimality criterion. A sequence $r_{n}$ is a minimax rate of testing
[see \citet{ingster:2000}] if the following statements are satisfied:
\begin{longlist}[1.]
\item[1.] For any $r'_n$ such that $r'_n/r_n \to0$ as $n\to\infty$,
%
\begin{equation}
\label{eq:rate-optimal-1} \liminf_{n\goto\infty}\inf
_{T} \bigl\{\alpha_n(T) + \beta
_n(T,1) \bigr\} = 1,
\end{equation}
where the infimum is taken on all decision rules, that is, $\{0,1\}
$-valued measurable functions of the sample $(X_1,\ldots,X_n)$.
\item[2.] For any $\alpha$, $\beta>0$, there exist some constant $C>0$ and
a test statistic $T^*$ (said \emph{rate optimal in the minimax sense}),
such that
%
\begin{equation}
\label{eq:rate-optimal-2} \limsup_{n\goto\infty} \alpha_n
\bigl(T^*\bigr) \leq \alpha\quad\mbox{and}\quad \limsup_{n\goto\infty}
\beta_n\bigl(T^*,C\bigr) \le\beta.
\end{equation}
\end{longlist}

Condition (\ref{eq:rate-optimal-1}) says that if the separation rate
vanishes faster that $r_n$, then no test can do better than the blind
random decision, for which the sum of the errors of the two kinds is
exactly 1. Condition (\ref{eq:rate-optimal-2}) says that there exists a
decision that is efficient for such a separation rate, so that this
rate is indeed a critical rate.

It is clear that a good test become sensitive to a closer and closer
alternative hypothesis $\hyp{1,n}$ when the sample size $n$ grows. The
notation of critical radius gives a precise and quantitative description of this
behaviour. The rate $r_n = 1/\sqrt{n}$ is the usual rate in the regular
parametric setting.

\subsubsection{Invariance properties}
\label{sec:invar-prop}

As the uniform distribution is invariant under rotations of the
sphere, the
theory of invariant tests [see \citet{lehmann:romano:2005}, Chapter 6]
leads to impose the same kind of invariance on any
statistical procedure for testing isotropy [see,
e.g., \citet{gine:1975} and the references therein].
As bases of invariant subspaces under rotations, the spherical
harmonics are thus the most natural tools to detect some deviation from
isotropy as in problem~(\ref{eq:T1}).
However, as explained earlier, a common property of astrophysical
observation of (point or continuous) processes on the sphere is the
nonuniform coverage of the \emph{sky} by the instrument. It is common
also that some parts of the data are missing or so noisy that it is
preferable to completely ignore or mask them.
That is why noninvariant approaches must be considered, and localized
analysis functions (such as wavelets) may be used as alternatives to
spherical harmonics. 
In the same spirit, wavelets have been proposed in the context of the
angular power spectrum estimation by \citet{baldi:etal:2009a} and used
in the realistic case of a partially observed stationary process with
heteroscedastic noise in \citet{fay:etal:2008} and \citet
{fay:guilloux:2011}.

\subsubsection{Regularity conditions: Besov spaces on the sphere}
\label{sec:regul-cond-besov}

Although this is not directly the purpose of this paper, it is a
natural question to ask which kind of regularity spaces our procedures
are designed for. The problem of choosing appropriate spaces of
regularity on the sphere is a serious question, and it is important to
consider the spaces which generalize usual approximation properties. On
the other hand, we are interested in spaces of functions which can be
characterized by their needlet coefficients $\{\beta\jk\}$ associated
to a needlet frame $\{\psi\jk\}$ (where $j$ denotes the scale and $k$
the position; see the online supplement [\citet{fay:etal:2013b}] for the
precise definitions). Hence, as is standard in the nonparametric
literature, it is natural to consider Besov bodies constructed on the
needlet basis. In many situations (not only the sphere) it can be
proved that these spaces can also be described as approximation spaces,
so they have a genuine meaning and can be compared to Sobolev spaces.
%
We define here the Besov body $B^s_{pq}$ as the space of functions
$f=(4\pi)^{-1}\int_\Sset f \dd\mu+ \sum_{j\ge0} \sum_{k\in\cK
_j }
\beta_{j,k} \psi_{j,k}$ such that
\[
\sum_{j \geq0}2^{jsq} \biggl(\sum
_{k\in
\cK_j } \bigl(|\beta_{j,k}| \|\psi_{j,k}
\|_p\bigr)^p \biggr)^{q/p} <\infty
\]
(with the obvious modifications for the cases $p$ or $q=\infty$).
%
Details on Besov spaces and their characterization by wavelets can be
found in \citet{triebel:1992} and~\citet{meyer:1992}. For details on the
relations between needlets and Besov spaces we refer, for instance, to
\citeauthor{narcowichetal:2006b} (\citeyear{narcowichetal:2006b,narcowich:petrushev:ward:2006}), \citet
{petrushev:xu:2008}.

\section{Needlet based test procedure and other anisotropy tests}
\label{sec:anisotropy-tests}

We introduce here two anisotropy detection procedures based on the
needlet analysis of $\{X_i\}_{i=1,\ldots,n}$. The first one is based on
multiple testing and will be referred to as \multiple. The second one
uses an estimate of the density plugged in a distance criterion and
will be referred to as \plugin. For the sake of further comparison (see
Section \ref{sec:monte-carlo-exper}), we also describe two existing
methods that are used in the gamma ray burst and cosmic ray literature.
The first one is based on a nearest neighbour analysis [see \citet{quashnock:lamb:1993b},
\citet{efron:petrosian:1995}]. The second one
relies on the two-point correlation [see, e.g., \citet{1993MNRAS.265L..65N},
\citet{kachelriess:semikoz:2006}].

We want detection procedures that are efficient from a $\Lset^2$ point
of view, but also for other $\Lset^p$ norms. In addition, we will
require procedures that are simple to implement as well as adaptive to
unknown and inhomogeneous smoothness. In Euclidean frameworks, these
types of requirements are well known to be efficiently handled by
(nonlinear) wavelet thresholding estimation in the context of density
estimation [see, e.g., \citet{donoho:etal:1996}] or by
multiple tests [\citet{ingster:2000},
\citet{spokoiny:1996}].

Our problem here requires a special construction adapted to the sphere,
since usual tensorized wavelets will never reflect the manifold
structure of the sphere and will necessarily create unwanted artifacts.
Recently a tight frame (i.e., a redundant family sharing some
properties with orthonormal bases), called a \emph{needlet frame}, was
produced which enjoys enough properties to be successfully used for
density estimation [\citet{baldi:etal:2009c}], for example, concentration
in the ``Fourier'' domain as well as in the space domain. Here,
obviously the ``space'' domain is the two-dimensional sphere itself,
whereas the Fourier domain is now obtained by replacing the ``Fourier''
basis by the basis of Spherical Harmonics which leads, as mentioned in
the previous section, to invariant tests. This construction produces a
family of functions $\{\psi\jk, j\ge0, k \in\cK_j\}$
which very much resemble wavelets. The index $k$ defines (with an
analogy to the standard wavelets) the \emph{locations} (points) on the
sphere around which the needlet is concentrated, and $j$ is referred to
as the \emph{scale}. These needlets have been shown to be extremely
useful for solving several types of astrophysical problems [\citet{delabrouille:etal:2008},
\citet{fay:etal:2008},
\citet{pietrobon:balbi:marinucci:2006},
\citet{marinuccietal:2008},
\citet{2008PhRvD..78j3504P},
\citet{2009ApJ...701..369R}]
or diverse inverse problems in statistics [\citet{kerkyacharian:etal:2007},
\citet{kerkyacharian:phamngoc:picard:2009},
\citet{kerkyacharian:etal:2009b}]. They are
especially well adapted to the situation recurrent in astrophysics
where the ``full sky'' is not covered (meaning in our context that
there are regions of the sphere where the points $X_i$ are not observed
if they happen to fall there).

A formal definition of needlets on the sphere is proposed in the
online supplement to this article [\citet{fay:etal:2013b}]
and can be found in greater detail in \citet
{narcowich:petrushev:ward:2006}. For the description of the test
procedures, we only need to define the empirical needlet coefficients
%
\begin{equation}
\hat\beta\jk\eqdef \frac{1}{n}\sum_{i=1}^n
\psi\jk(X_i), \label{eq:defhatbeta}
\end{equation}
which are unbiased estimators of $\beta\jk(f) \eqdef\langle f, \psi
\jk
\rangle= \int_\Sset f(\xi)\psi\jk(\xi) \dd\xi$. As usual in the
wavelet literature, $j \geq0$ refers to the scale and $k$ to the
location. The coarsest scale is $j=0$. The index $k$ refers to a
collection of quadrature points $\{\xi_{j,k}\}$ that are available at
each scale $j$. $\psi_{j,k}$ is then a zero-mean function centered on
$\xi_{j,k}$ and more and more concentrated as $j \to\infty$.

\begin{figure}

\includegraphics{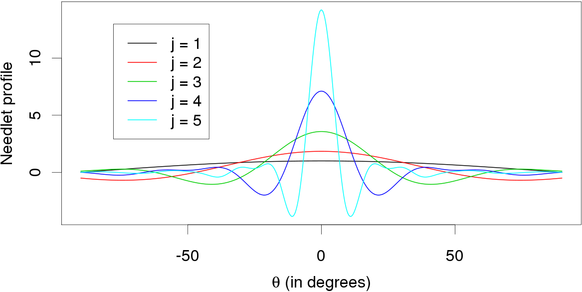}

\caption{The shape of the five first needlets in the spatial domain as
the function of the co-latitude~$\theta$. Recall that all the $\psi
_{j,k}$ functions are axisymmetric around the points $\xi_{j,k}$.}
\label{fig:needlets}
\end{figure}

In our simulations, we have chosen dyadic needlets with a spline
function of order 15 as generator, which leads to simple but
sufficiently concentrated analysis wavelets.
All the wavelets are axisymmetric around some well chosen points $\xi
_{j,k}$. The spatial profiles of those needlets at the five coarsest
scales are represented in Figure \ref{fig:needlets}. More details are
available in the online supplement [\citet{fay:etal:2013b}].

\subsection{Multiple tests}
\label{sec:multiple-tests}

For multiple tests, we will consider collections of ``linear
estimators'' of the density, meaning that we will not use any nonlinear
processing of wavelet coefficients such as thresholding in the
estimation phase. By analogy with the work of \citet
{butucea:tribouley:2006} on the related problem of the two-sample
nonparametric homogeneity test, we define
%
\begin{equation}
\hat f_J = \frac{1}{4\pi} + \sum_{j = 0}^{J}
\sum_{k \in\cK_j} \hat\beta\jk\psi\jk\label{eq:deflinest}
\end{equation}
with the $\beta\jk$'s given by (\ref{eq:defhatbeta}).
For any value of the smoothing parameter $J$, we define the
nonrandomized associated testing procedure
%
\begin{equation}
\label{eq:defTjlinear} T_J = \ind_{d(\hat f_J,g)\ge t_{J}} = %
\cases{1,&  \quad$\mbox{if } d(\hat f_J,g) \ge t_{J}$,
\vspace*{2pt}
\cr
0, & \quad$\mbox{if } d(\hat f_J,g) < t_{J}$.}
\end{equation}
This gives a family of tests indexed by $J$, where the dependence with
respect to the choice of the distance $d$ and to the sequence of
thresholds $t_J$ is made implicit in the notation.

\citet{butucea:tribouley:2006} proved that if the regularity conditions
are known and specified by Besov conditions, the smoothing parameter
$J$ can be chosen optimally. It is likely that their arguments could be
reproduced in our case. However, our point of view in this paper will
not be to detail this theoretical issue but rather to concentrate on
the practical aspects of the tests. Moreover, it would be probably
difficult to relate physical information to mathematical regularity conditions.

Nevertheless, the optimal choice for the parameter $J$ depends on the
regularity~$s$ specified in the class of alternatives. Adaptive
optimality can be achieved thanks to a multiple test that decides for
the alternative hypothesis as soon as one of the $T_J(d,c_J)=1$
individually does so, that is, defining $T^{\scriptsize\multiple}$, by
%
\begin{equation}
T^{\scriptsize \multiple}= 0\quad \mbox{if and only if}\quad \forall{J\le{J^{\star}}},
T_J = 0.
\end{equation}
Mimicking the theoretical results obtained in \citet
{butucea:tribouley:2006} and \citet{baldi:etal:2009c}, we have used
%
\begin{equation}
{J^{\star}}= \bigl\lfloor\tfrac12 \log_2(n/\log n) \bigr
\rfloor \label{eq:expressionbandmax}
\end{equation}
as a reference in our numerical investigations, as in the case of
adaptive density estimation (see below). Note, however, that the
optimal ${J^{\star}}$ could vary according to the loss
function ($\Lset^p$
norm) we use to measure the nonisotropy as suggested by the results of
the related problem in the two sample nonhomogeneity detection \citet
{butucea:tribouley:2006}. The values $t_J$ that are used in (\ref
{eq:defTjlinear}) must be chosen to verify $\prob_{g}(T^{\scriptsize\multiple}=0)
\simeq\alpha,$ where $\alpha$ is the prescribed level of the test.
\subsection{Plug-in tests}
\label{sec:plugin-tests}

It is also interesting to compare, from an empirical point of view, the
above multiple test procedures to algorithms where we simply plug in an
adaptive estimate of the density in the distance. These density
estimators have good asymptotic properties from a minimax point of
view, hence, it makes sense to investigate also their properties when
used for testing. To the best of our knowledge, no theoretical
optimality is proved and there even are arguments suggesting that these
procedures might not be optimal. For instance, on the real line, the
minimax rate of convergence for estimation (in the so-called dense
case) is $n^{-s/(2s+1)}$, meaning that if $f$ belongs to a ball in a H\"
{o}lder space with exponent $s$, then no estimator can approach the
least favorable density at a better error rate (measured in a $\Lset^p$
norm). We refer to \citeauthor{donoho:etal:1996} [(\citeyear{donoho:etal:1996}), Theorem 3] for a precise
statement, among others.
On the other hand, the minimax critical radius for nonuniformity
detection is $n^{-2s/(4s+1)}$ [see \citet{ingster:2000}]. It means
that, in the minimax framework, one can distinguish asymptotically two
hypotheses that are separated by a distance negligible with respect to
the accuracy of any nonparametric estimation of the densities in an
infinite dimensional space.

The most popular minimax adaptive technique consists in adding to a
very basic linear estimation a thresholding rule as post-processing. In
the above mentioned paper [\citet{baldi:etal:2009c}] this nonlinear
post-processing actually is a \emph{hard} thresholding rule, namely,
\[
\hat f_{{J^{\star}}}= \frac{1}{4\pi} + \sum_{j =
0}^{{J^{\star}}}
\sum_{k
\in\cK_j} \hat\beta\jk\ind_{|\hat\beta\jk| > \kappa\sqrt
{\log n/n}} \psi\jk
\]
for some positive constants $\kappa$ and ${J^{\star}}=
\lfloor\tfrac12
\log_2(n/\log n) \rfloor$.
The coefficients $\hat\beta\jk$ are defined in (\ref{eq:defhatbeta}).

It is known that many variations exist with close theoretical
properties but some differences in different practical situations.
Among those, we will especially consider the data-driven thresholding
introduced by \citet{juditsky:lambert-lacroix:2004} to deal with density
estimation on the real line (as opposed to density on $[0,1]$). It seems
to give good detection procedures for small samples in our context. In
the following, we will consider the nonlinear estimates
%
\begin{equation}
\label{eq:defhatf} \hat f_{{J^{\star}}} = \frac
{1}{4\pi} + \sum
_{j =
1}^{{J^{\star}}}\sum_{k \in\cK_j}
\ind_{|\hat\beta\jk| > \lambda\sqrt{\log n}\hat\sigma\jk} \ind _{\delta
\jk>\rho\log n} \hat\beta\jk\psi\jk
\end{equation}
for some positive constants $\rho, \lambda$, ${J^{\star
}}= \lfloor
\tfrac
12 \log_2(n/\rho\log n) \rfloor$, and where
%
\begin{eqnarray}
\hat\sigma\jk^2 &\eqdef&\frac1 n\sum_{i=1}^n
\psi\jk^2(X_i) - (\hat \beta\jk)^2,
\label{eq:defsigmajk}
\\
\delta\jk&\eqdef&\bigl(\psi\jk^2(\xi\jk)\bigr)^{-1} \sum
_{i=1}^n \psi\jk^2(X_i). \label{eq:defdeltajk}
\end{eqnarray}

Let us give a short interpretation of the thresholding procedure. The
quantity $\hat\sigma\jk^2$ is an estimate of the variance of $\hat
\beta\jk$. The expression for $\delta\jk$ is inspired by the one
provided in \citet{juditsky:lambert-lacroix:2004}. In this reference,
compactly supported wavelets on the real line are used with a threshold
on the number of observations actually participating to the estimation
of $\beta_{jk}$. In this case, it makes sense to count the number of
observations falling in the support of the wavelet. In our case, as
needlets are supported on the whole sphere (although very
concentrated), we propose to replace this quantity by a continuous type
approximation~$\delta\jk$; see (\ref{eq:defdeltajk}). Note that
$\delta
\jk= n$ if $X_1=\cdots=X_n=\xi\jk$.

Finally, we define the \plugin{} procedure as the decision
%
\begin{equation}
\label{eq:defTjplugin} T^{\scriptsize\plugin{}}_J = \ind_{d(\hat f_{J^*},g) \ge t^{\scriptsize\plugin{}}_{J}},
\end{equation}
with $\hat f_{{J^{\star}}}$ defined in (\ref
{eq:defhatf}) and $t^{\scriptsize\plugin}
_{J}$ some fixed threshold depending on the prescribed level $\alpha$
of the test.

\subsection{Two-point correlation test and nearest neighbour test}
\label{sec:two-point-corr}

When dealing with one-dimensional data, one can compare every test
procedure to the well-known benchmark Kolmogorov--Smirnov or Cram\'
{e}r--von Mises tests, which are based on the empirical distribution
function of the sample. In higher dimensions (here on the sphere),
there is no natural order relation that allows to consider such
approaches. For sake of comparison, we have run some simulations on two
different tests found in the astronomical literature.

\subsubsection*{Nearest neighbour test} The following statistical
procedure has been proposed by \citet{quashnock:lamb:1993b}. We denote it
\efron, as nearest neighbour. For each point $X_i$, we compute the
distance $Y_i$ to its nearest neighbour. Under the hypothesis that $f$
is uniform over the whole sphere, the marginal distribution function of
$(Y_i)$ is $\phi\dvtx  y \mapsto1 - [(1+\cos y)/2]^{n-1}$, and the Wilcoxon
statistic
\[
W = \sqrt{12 n} \Biggl( \frac12 - \frac1n \sum_{i=1}^n
\phi(Y_i) \Biggr)
\]
is asymptotically standard Gaussian.
For a nonhomogeneous random draw (for instance, in the presence of
clusters), this
statistic is expected to take significantly high values, allowing to
detect this kind of anisotropy.
This test is of interest, as it is simple to compute, it has no
parameters to be tuned, and it admits an
extension to nonuniform exposure [see \citet{efron:petrosian:1995}].
In this case, the distribution of $W$ is estimated numerically by Monte
Carlo methods.
The \efron{} procedure simply writes
%
\begin{equation}
\label{eq:2} T^{\scriptsize\efron}= \ind_{W \geq t^{\scriptsize\efron}},
\end{equation}
where $t^{\scriptsize\efron}_{1-\alpha}$ is the $(1-\alpha)$-quantile of the
distribution of $W$.
This distribution can be approximated by a standard Gaussian
distribution if the sample size is big and the exposure is uniform.
Otherwise, the quantile is estimated by the Monte Carlo method.

\subsubsection*{Two-point correlation test} Among others, \citet
{1993MNRAS.265L..65N,kachelriess:semikoz:2006} use the empirical
two-point autocorrelation function to detect clustering (\twopc{}
test). For a collection of $n$ points $\{X_i\}$ and any angular
distance $\delta\in[0,\pi]$, let $N_n(\delta)$ denote the random
number of pairs $\{i,j\}$ such that $\Delta(X_i,X_j) \leq\delta$,
where $\Delta$ is the geodesic distance. Define the two-point
correlation function $w_n(\delta)= \esp(N_n(\delta))$ and its empirical
counterpart
%
\begin{equation}
\label{eq:4} \hat w_n(\delta) = \sum_{i < j}
\ind_{[0, \delta]}\bigl(\Delta (X_i,X_j)\bigr).
\end{equation}
Under the null hypothesis, the distribution of $\hat w_n$ at any
$\delta
_0$ is evaluated using Monte Carlo simulations. Then, the detection
will be based on the comparison between the empirical correlation
function and $w_n$, at some fixed value $\delta_0$ or a few different
values. A typical $\delta_0$ can be chosen so as to maximize the
sensitivity of the test depending on the application. In some
references, however, the probability to observe a value bigger than
$\hat w_n(\delta)$ is plotted on the whole range $[0,\pi]$ with no
$\delta_0$ fixed a priori. Consequently, much care is taken when
interpreting those values, as stressed, for instance, in \citet
{kachelriess:semikoz:2006}.
Here we define the procedure \twopc{} by the decision
%
\begin{equation}
\label{eq:3} T^{\scriptsize\twopc}= \ind_{\hat w_n(\delta_0) \ge t^{\scriptsize\twopc}},
\end{equation}
where $t_{1-\alpha}^{\scriptsize\twopc}$ is the $(1-\alpha)$ quantile of the
distribution of $\hat w_n(\delta_0)$ under the null, evaluated by Monte
Carlo simulations, at some $\delta_0$ specified a priori.

\section{Monte Carlo experiments}
\label{sec:monte-carlo-exper}

\subsection{Experimental setup}
\label{sec:experimental-setup}

In this section we compare numerically the tests defined in
Section \ref
{sec:anisotropy-tests} that are denoted \multiple{}, \plugin{},
\efron{} and \twopc{}.

For $T$ being any of those nonrandomized test procedures, we can tune
the parameters of the procedure to have a prescribed level $\alpha$,
that is,
$\prob_g(T = 1) = \alpha$. This is done by Monte Carlo replication. Ten
thousand independent random samples of size $n$ are drawn under the
null hypothesis, for $g$ being the uniform density on $\Sset$ [\ie, $g
\equiv1/(4\pi)$] or the stylized exposure function of the Pierre Auger
detector (see Figure \ref{fig:auger_exposure}).

For the \multiple{} procedure and a given level $\alpha$, we have chosen
%
\begin{equation}
\label{eq:defTj} T_{j} = \ind_{\|\hat f_j - g\|_p > t_{\alpha',j}},
\end{equation}
where $t_{\alpha',j}$ is the $1-\alpha'$ quantile of the distribution
of $\|\hat f_j - g\|_p$ under the null hypothesis. This distribution is
evaluated using Monte Carlo replications. Further, the value $\alpha'$
is chosen so that
%
\begin{equation}
\label{eq:defUTj} T'_{{J^{\star}}} = \sup_{j=1,\ldots,{J^{\star}}}
T_j
\end{equation}
has a first type error probability equal to $\alpha$. This is arbitrary
and the theory to be written would likely suggest to use a scale
dependent level.

\def\bandmax{{J^{\star}}}
\def\plugintxt{\textsc{PlugIn}}
\def\multipletxt{\textsc{Multiple}}
\def\efron{\textsc{NN}}
\def\twopc{\textsc{TwoPC}}

\begin{table}
\caption{Power (in \%) under $\hypalt{c}$, under uniform exposure, with
$n_s=100$ and $E_{\min} = 10^{19}$~\textup{eV}}\label{tab:H3power1}
\begin{tabular*}{\textwidth}{@{\extracolsep{\fill}}lccccccccc@{}}
\hline
&&\multicolumn{4}{c}{$\bolds{n=25}$}& \multicolumn{4}{c}{$\bolds{n=100}$}\\[-6pt]
&&\multicolumn{4}{c}{\hrulefill}& \multicolumn{4}{c@{}}{\hrulefill}\\
&  \multicolumn{1}{c}{$\bolds{\bandmax}$}& \textbf{3}& \textbf{4}& \textbf{5}& \textbf{6}&  \textbf{3}& \textbf{4}& \textbf{5}&\textbf{6}\\
\hline
\textsc{Multiple}& $p=1$ & 51 & 46 & 41 & 40  &98 & 98 & 98 & 98\\
& $p=2^\star$ & 52 & \textbf{53} & 47 & 47 & 98 & \textbf{99} & 98 & 98\\
& $p=\infty$ & 42 & 44 & 42 & 42 & 92 & 91 & 91 & 90\\[3pt]
\textsc{PlugIn}& $p=1$ & 34 & 34 & 34 & 34 & 98 & 98 & 98 & 98\\
& $p=2$ & 42 & 42 & 42 & 42 & 98 &  98 &  98 & 98\\
& $p=\infty$ & 50 & 50 & 50 & 50& 92 & 92 & 92 & 92\\[3pt]
\textsc{NN}&& \multicolumn{4}{c}{38}& \multicolumn{4}{c}{82}\\
\textsc{TwoPC}&& \multicolumn{4}{c}{45} &\multicolumn{4}{c}{62} \\
\hline
\end{tabular*}
\end{table}

\begin{table}
\caption{Power (in \%) under $\hypalt{c}$, under uniform exposure, with
$n_s=500$ and $E_{\min} = 6\times 10^{19}$~\textup{eV}}\label{tab:H3power3}
\begin{tabular*}{\textwidth}{@{\extracolsep{\fill}}lccccccccc@{}}
\hline
&&\multicolumn{4}{c}{$\bolds{n=25}$}& \multicolumn{4}{c}{$\bolds{n=100}$}\\[-6pt]
&&\multicolumn{4}{c}{\hrulefill}& \multicolumn{4}{c@{}}{\hrulefill}\\
&  \multicolumn{1}{c}{$\bolds{\bandmax}$}& \textbf{3}& \textbf{4}& \textbf{5}& \textbf{6}&  \textbf{3}& \textbf{4}& \textbf{5}&\textbf{6}\\
\hline
\textsc{Multiple} & $p=1$ & 27 & 39 & 45 & 43 & 76 & 94 & \phantom{0}99 & \phantom{0}98\\
& $p=2^\star$ & 28 & 42 & 50 & 58& 79 & 96 & 100 & \textbf{100} \\
& $p=\infty$ & 24 & 35 & 45 & 50 &69 & 84 & \phantom{0}96 & \phantom{0}97\\[3pt]
\textsc{PlugIn} & $p=1$ & 18 & 18 & 18 & 18 &72 & 72 & \phantom{0}73 & \phantom{0}73\\
& $p=2$ & 25 & 28 & 29 & 29 &78 & 82 & \phantom{0}82 & \phantom{0}82\\
& $p=\infty$ & 33 & 34 & 34 & 34 &71 & 80 & \phantom{0}80 & \phantom{0}80 \\[3pt]
\textsc{NN}&& \multicolumn{4}{c}{33}& \multicolumn{4}{c}{99}\\
\textsc{TwoPC}&& \multicolumn{4}{c}{\textbf{75}} &\multicolumn{4}{c}{99} \\
\hline
\end{tabular*}
\end{table}

The power of the test $T$ is defined by (\ref{eq:defpowertest}). Some
clues about this value are obtained by evaluating $\prob_f(T=1)$ for
particular alternatives $f$ that are given in the next section. Here
again, those quantities are evaluated by Monte Carlo. Note, however,
that the power for a particular alternative only gives an upper bound
of the power in the minimax sense given by the second equation of (\ref
{eq:defpowertest}).

In the tables of tests in the main paper (Tables \ref{tab:H3power1} through
\ref{tab:tablerate}) and on its online supplement [Tables 1 through 8
in \citet{fay:etal:2013b}], 
we represent the power of four tests. The power of the needlet tests
is expressed as a function of the finest band $J^{\star}$ and the power of
the norm we use to detect anisotropy (see the online supplement [\citet
{fay:etal:2013b}]
for more details on the actual implementation of the method).

The profile cuts of the (axisymmetric) needlets we have used are
plotted in the online supplement.

\begin{table}[b]
\caption{Power of the tests for three models of $\hypalt{a}$ with values
of $\delta$ and sample size varying so that $\sqrt{n}d(f,g)$ remains
constant. It appears that those particular sequences of powers are
generally nondecreasing with the sample size. The observation model
uses the Pierre Auger exposure function}\label{tab:tablerate}
\begin{tabular*}{\textwidth}{@{\extracolsep{\fill}}lccccccccccccc@{}}
\hline
&
&
\multicolumn{4}{c}{$\bolds{n=25}$\textbf{,} $\bolds{\delta=0.0}8$}&
\multicolumn{4}{c}{$\bolds{n=100}$\textbf{,} $\bolds{\delta=0.04}$}&
\multicolumn{4}{c}{$\bolds{n=400}$\textbf{,} $\bolds{\delta=0.02}$}\\[-6pt]
&
&
\multicolumn{4}{c}{\hrulefill}&
\multicolumn{4}{c}{\hrulefill}&
\multicolumn{4}{c@{}}{\hrulefill}\\
& $\bolds{\bandmax}$& \textbf{3}& \textbf{4}& \textbf{5}& \textbf{6} & \textbf{3}& \textbf{4}& \textbf{5}& \textbf{6}
& \textbf{3}& \textbf{4}& \textbf{5}& \textbf{6}\\
\hline
\textsc{Multiple}& $p=1$ & 14 & 16 & 13 & 13
& 14 & 16 & 14 & 14
& 21 & 20 & 17 & 17 \\
& $p=2^\star$ & 19 & 20 & 16 & 16
& 17 & 21 & 20 & 20
& 23 & 21 & 20 & 20 \\
& $p=\infty$ & 23 & {26} & 23 & 22
& 29 & {32} & 32 & 30
& {34} & 32 & 30 & 29 \\[3pt]
\textsc{PlugIn}& $p=1$ & 11 & 11 & 11 & 11
& 17 & 16 & 16 & 16
&19 & 19 & 19 & 19 \\
& $p=2$ & 16 & 16 & 16 & 16
& 26 & 27 & 27 & 27
& 32 & 32 & 32 & 32 \\
& $p=\infty$
& {23} & 22 & 22 & 22
& \textbf{32} & 30 & 30 & 30
& \textbf{39} & \textbf{39} & \textbf{39} &
\textbf{39} \\[3pt]
\textsc{NN}&& \multicolumn{4}{c}{\phantom{0}8} & \multicolumn{4}{c}{\phantom{0}6}
& \multicolumn{4}{c@{}}{\phantom{0}5} \\
\textsc{TwoPC}&& \multicolumn{4}{c}{\textbf{35}} & \multicolumn
{4}{c}{14} &
\multicolumn{4}{c@{}}{14} \\
\hline
\end{tabular*}
\end{table}

\subsection{Alternatives}
\label{sec:alternatives}

We have investigated the performance of the test (power against level)
for sample sets of small to moderate size ($n=25, 100, 400$) and
against different alternatives. Those choices of $n$ mimic the
progressive publication of events by the Pierre Auger Observatory (27
events above $5.7 \times10^{19}$~eV in 2008, 69 above $5.5 \times
10^{19}$ in 2010, a few hundred in the future).

Generally speaking, the physical plausibility of those alternatives is
weak [alternative $\hypalt{c}$], if not null [alternatives $\hypalt
{b}$ and
$\hypalt{c}$]. Our goal is to focus here on specific departures from
isotropy. First we consider unimodal nonisotropic densities, with a
Gaussian shape. Then we consider mixtures of densities that would only
be obtained if the sources of the cosmic rays were known to be
uniformly distributed and repeating, and at the same distance from us.
Third, the Physics-inspired model $\hypalt{c}$ gives rise to nonisotropic
patterns with richer frequency content compared to the previous ones
(and nonaxisymmetric clusters). We now give the precise definitions of
the alternatives.

\subsubsection*{$\hypalt{a}$} 
\begin{figure}

\includegraphics{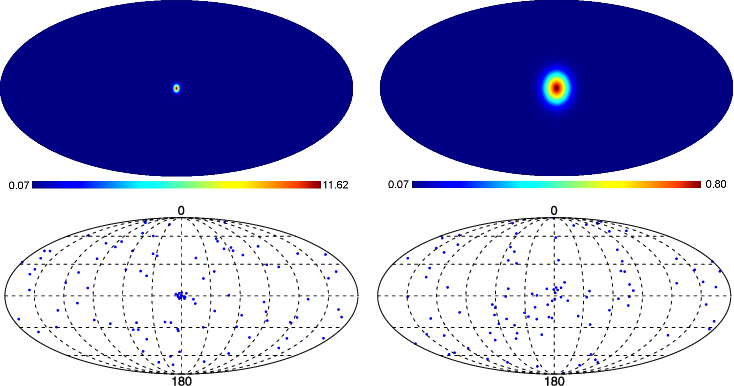}

\caption{Densities (first line) and random draws (second line,
$n=100$) under $\hypalt{a}$ with $\delta= 10\%$ and $\theta= 5^\circ$
(left) or $\theta= 20^\circ$ (right).}
\label{fig:H1exemple}
\end{figure}

The first family of alternatives is obtained as
a mixture of the uniform density $h_0$ and an over-density at some
point of the sphere, with Gaussian-like axisymmetric profile.
Precisely, the density under $\hypalt{a}$ writes
\[
h(\xi) = (1-\delta)h_0 + \delta h_\theta(\xi),
\]
where for $\theta\neq0$, we put $h_\theta(\xi):= h_{\theta,\xi
_0}(\xi
)$, $h_{\theta,\xi_0}:= C_\theta\exp(-(\xi\cdot\xi_0)^2/2\theta^2)$
and $\xi_0 = (\pi/2,0)$. Such densities are then unimodal, with a bump
whose width is proportional to $\theta$.
Typical observations of random draw with such density with $\delta
=0.01$ and $\theta=5^\circ$ or $20^\circ$ are displayed on
Figure \ref
{fig:H1exemple}.
%

%

\subsubsection*{$\hypalt{b}$} 
\begin{figure}

\includegraphics{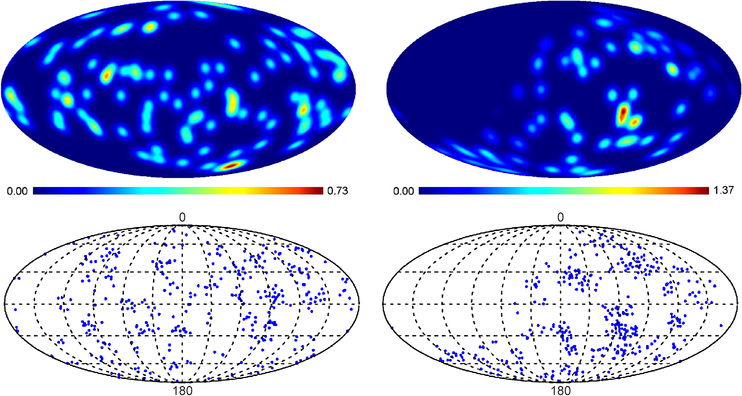}

\caption{Density of $X_1$ conditionally to the random draw of the
centers of 100 AGNs (first line) and random draws with $n=400$ (second
line). The exposure is uniform on the left, \emph{\`{a} la} Pierre
Auger Observatory on the right.}
\label{fig:H2exemple} 
\end{figure}

A second family of alternatives is a toy model
for the repeating emission of events from a small number of sources, as
explained in the \hyperref[sec:introduction]{Introduction}. Here we assume that the $n_s$ sources
are uniformly distributed, although in a realistic case, we can expect
any type of astrophysical sources to follow the local matter density of
the cosmic structure (which would make the detection of anisotropy
easier). This generalization is straightforward enough that we do not
discuss it further at this stage. Conditionally to those positions, the
incidental directions are distributed along a mixture of $n_s$ Gaussian
densities centred on the sources (to take into account the error in the
measurement of the incidence angle or the deflection of the charged
particle by Galactic magnetic fields), namely,
\[
h(\xi) = \sum_{j=1}^{n_s}
h_{\theta,\xi_i}(\xi).
\]
This density is then modulated by the exposure $\varepsilon$ of the
detector along equation (\ref{eq:density_observations}). Such
conditional densities are displayed on the first line of Figure \ref
{fig:H2exemple} with uniform and Pierre Auger exposures. We considered
the cases $n_s=10$ and $n_s=100$ and fixed $\theta= 10^\circ$. Note
that if $n_s$ is much bigger than $n$, it is difficult to detect this
kind of anisotropy (which can be detected only if at least one source
has emitted more than one cosmic ray).
%

%

\subsubsection*{$\hypalt{c}$}
\begin{figure}

\includegraphics{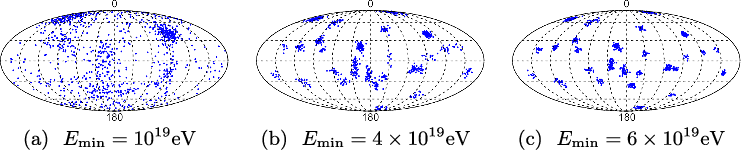}

\caption{Isotropization of the cosmic rays in model $\hypalt{c}$ as
$E_{\min}$ decreases. There are the exact same $n_s = 30$ sources in
the three cases and $n=1000$ observations.}
\label{fig:H3example}
\end{figure}

A third and last alternative is obtained by the physical model of
cosmic ray observations described in detail in Section \ref{toymodel}.
Sources are randomly drawn in a spherical volume of radius $r_{\max} =
70$ Mpc, and their flux is assumed inversely proportional to the square
of their distance.
The parameters for the simulations are taken to be $E_{\max} =
10^{21}$~eV, $\alpha= 4.2$. We consider different values for $E_{\min}$
(namely, $1$, $4$ or $6$ $\times10^{19}$~eV). Playing on this parameter
has an important practical incidence. Assuming that the distribution of
the energy of the cosmic rays is a power law, $\prob(E > t) \sim C
t^{-\alpha+ 1}$, lowering the threshold on the selection of the cosmic
rays from $6 \times10^{19}$~eV to $4 \times10^{19}$~eV (resp.,
$10^{19}$~eV) accounts to increase the size of the sample (available
observations above the threshold) by a factor $(6/4)^{\alpha-1} \simeq
3.66$ (resp., 310). It means that the statistical decision should be made
far easier if the cosmic rays were not too much isotropized by the
Galactic fields as their energies go lower. This effect is illustrated
in Figure \ref{fig:H3example}. It is interesting to see if the methods
are still able to detect anisotropy as the cosmic rays become more and
more isotropized.
%
%
%
%
This is a more realistic simulation compared to models $\hypalt{a}$ and
$\hypalt{b}$. There is no single size for the scatter of the CRs coming
for a given source, nor the same size or directionality for each
source, nor the same flux for each source, that hence is interesting
specifically for a multiscale analysis with no prior assumption about a
correlation length.

Note that under the alternatives $\hypalt{b}$ and $\hypalt{c}$, the
procedure is to be understood as a test on the conditional distribution
of $(X_i)_{i=1,\ldots,n}$ with respect to the positions of the
``sources'', which are randomly drawn once for all.

\subsection{Numerical results and discussion}
\label{sec:discussion-results}
\subsubsection*{Tables}
We shall represent some of the results of our simulations with tables of
estimated power of the procedures for given alternatives (in percent),
at the
prescribed level $\alpha= 0.05$. Practically, we let the finest
needlet band
entering the \multiple{} and \plugin{} procedures vary in the set
$\{{J^{\star}}-2,{J^{\star
}}-1,{J^{\star}},{J^{\star}}+1\}$ where
${J^{\star}}$
is given
by (\ref{eq:expressionbandmax}). The entry (or entries) corresponding
to the
overall highest power (before rounding off) among the 26 values is (are)
printed in bold type. We consider three $L^p$ norms, namely, $L^p$ for $p=1,2,
\infty$. It is possible to use an unbiased estimate of the distance between
$\hat f $ and $g$ in the case of the $L^2$ norm. It is referred to as
$p=2^\star$ (see the online supplement for details)

\subsubsection*{ROC curves}
The receiver operating characteristic (ROC) curves plot the power $p$
of a\vadjust{\goodbreak} procedure as a function of its level $\alpha$. It is a useful
representation for comparison of different procedures along a wide
range of levels. The ROC curves associated to the \twopc{} procedure
are a step function because of the discrete nature of the test
statistic. Some of the ROC curves are nonconcave. It should be
recalled, however, that any procedure of this kind can be improved to a
randomized procedure whose ROC curve is the concave upper envelope of
the original one. Accordingly, the reader's eyes must actually analyse
the upper envelopes of the ROC curves. Note that the power in the
tables has not been modified by this argument.

ROC curves are represented in plots with four subplots, corresponding
to the four above-mentioned choices of $J^{\star}$ in
the needlet
methods. The ROC curves for \twopc{} and \efron{} procedures are the
same in the four subplots.
Inset graphs allow complementary comparison of the methods by zooming
on the most relevant levels (small $\alpha$).

\subsubsection{Some specific results}
\label{sec:some-spec-results}

\def\plugin{\textsc{PlugIn}}
\def\multiple{\textsc{Multiple}}

First, we note that the differences of sensitivity between the
different $L^p$ norms we use are not very strong, probably because we
consider quite regular alternative hypotheses. As expected, the $\Lset
^\infty$ is a bit more sensitive to more spiky (unimodal)
distributions, whereas more global measures such as $\Lset^1$ or
$\Lset
^2$ perform better under the $\hypalt{b}$ or $\hypalt{c}$ models.
This is
illustrated by some ROC curves 
in the online supplement.
We now illustrate the comparison of the performances of the four
procedure with a few tables and figures.

\begin{figure}

\includegraphics{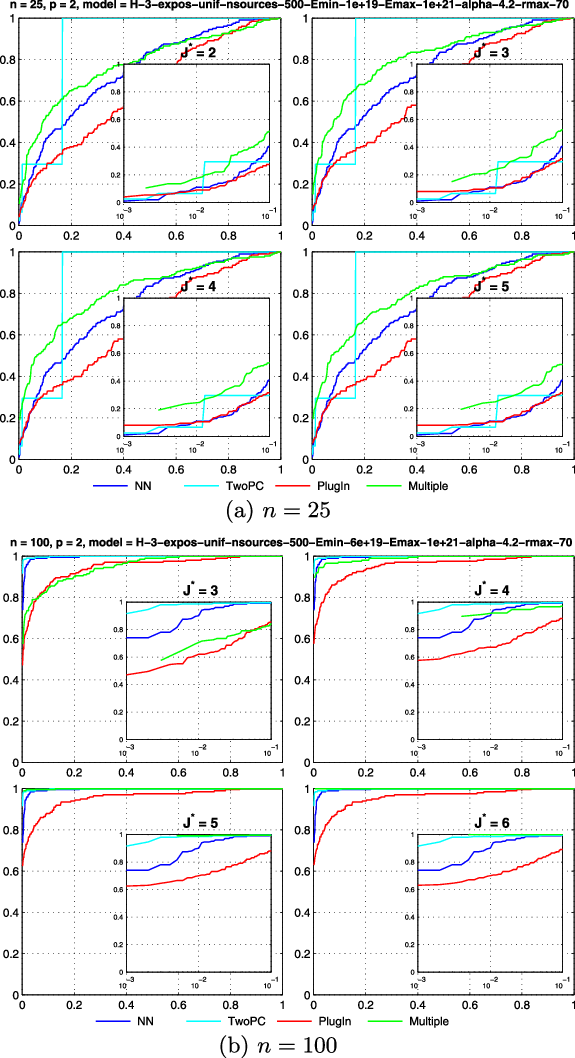}

\caption{
ROC curves (true positive rate against false positive rate) for the
four methods. For the needlet methods, the debiased $\Lset^2$ norm is
used. Insets display the same curves as in the main plot with a
logarithmic scale in abscissas, to highlight the comparative
performances for relevant level values.}
\label{fig:H3n25and100ns500}
\end{figure}

It appears that
the methods \multiple{} and \plugin{} have a consistent behaviour when
the typical radius of the anisotropic structure is varying. We shall
discuss further from those cases below. Figure \ref
{fig:H3n25and100ns500} illustrates their good performances even for
small samples under the model $\hypalt{c}$ that produce clusters of
various sizes and shapes.

%
%
%


The \efron{} procedure performs strikingly worse than others in almost
all but the $\hypalt{b}$ situations. The good sensitivity to $\hypalt
{b}$ alternatives can be explained in the following manner. In this
case, the points $\{X_i\}_{i=1,\ldots,n}$ are mainly grouped into
clusters of average scale given by the standard deviation of the
Gaussians of the mixture. If the number of clusters and this standard
deviation are too small to cover significantly the whole observed part
of the sphere, then the random distances to the nearest neighbour are
bounded by $\sigma$ with very high probability, which is not the case
under the null. This makes the distribution of the distance to the
nearest neighbour a very sensitive tool to discriminate between
$\hypalt
{b}$ and the null.

Varying the alternatives, it appears that no method outperforms the
other in a uniform way, but it seems that the two needlet methods, if
not always optimal, consistently have a good behaviour. Moreover, the
\multiple{} test is slightly more sensitive that the \plugin{} one. As
an illustration, we represent in Tables \ref{tab:H3power1} and \ref
{tab:H3power3} the power of the procedures against the $\hypalt{c}$
alternative, for sample sizes equal to 25 and 100, and $(E_{\min}, n_s)
= (10^{19}$~eV, 100), and ($6\times10^{19}$~eV, $500$), respectively. It
can be seen from those tables that moving the lower energy limit
upwards makes the detection easier.
More tables are available in the online supplement,
for a representative panel of alternatives, containing more or less
spiky distributions, clusters of smoother alternatives, weak or strong
anisotropy etc.

It must be stressed that the \twopc{} approach often provides a good
sensitivity if not the best at $n=25$. For most of the alternatives,
however, one or the other of the needlets methods outperforms \twopc{}
as $n$ grows. This is exemplified in Tables \ref{tab:H3power1} and 
\ref{tab:H3power3} in the case of a $\hypalt{c}$ alternative.
In our application context, the sample size over a given energy
threshold is increasing with time and experiments, so it must be
highlighted that multiscale methods are more and more appropriate for
analysis of future data sets.

\subsubsection{Separation rate}

We focus here on the behaviour of the power of the test with respect to
$n$. If $r_n$ is the critical rate in the minimax sense [given by
equations (\ref{eq:rate-optimal-1}) and (\ref{eq:rate-optimal-2})], we
should observe an approximately same power for different sample size
and the least favourable alternative densities $\tilde f_n$ as soon as
the quantity $r_n d(\tilde f_n,g)$ remains constant.
On Table \ref{tab:tablerate} we have displayed the power of the
different procedures for three different densities corresponding to the
alternative $\hypalt{a}$ and three sample sizes, keeping the same value
for $n^{1/2}d(f,g)$.
Indeed, in the $\hypalt{a}$ case, for any power norm, $d(h,h_0) =
\delta
d(h_0,h_\theta)$.
As the power remains roughly the same in $(0,1)$ for the three values
of the parameters, and as $n^{1/2}$ is an upper bound for the minimax
separation rate in analogy with similar problems on Euclidean spaces,
this numerical simulation is consistent with the claim that the needlet
based procedures perform well at the minimax rate of testing.
The increasing value of the power with $n$ together with the unbeatable
rate of separation $\sqrt{n}$ illustrates the fact that we only have
access to upper bounds of the minimax rate. In other words, the
densities under consideration are definitely not the least favourable cases.
The comparison of needlet methods with \efron{} and \twopc{} methods
tends to be more favourable to needlets methods as $n$ becomes larger
in this case.

\subsubsection{Robustness}

Assume that the anisotropy detection by the needlet methods is
adaptive. Then, as pre-tuned black boxes, those methods should remain
optimal on a wide range of alternatives. Some simulations support this
claim. Note, however, that we only explore physically possible
alternatives which are smooth nonuniform densities.

The key parameter of the \twopc{} method is the angular size $\delta_0$
at which we compare $\hat w(\delta)$ to the distribution of $w(\delta
_0)$ under the null. For sake of fairness in our comparisons, we should
allow some tuning of this parameter. It is clear that the optimal
$\delta_0$ is related to the ``average scale'' of the anisotropy.
Though it is difficult to give a precise and general definition of this
former quantity, it should be close to the value of the parameter
$\theta$ in the particular case of model $\hypalt{a}$. Indeed, it appears
from our simulations that \twopc{} is better than the needlet methods
when $\theta= 5^\circ$ and worse when $\theta= 20^\circ$ under
$\hypalt{a}$.

\begin{figure}

\includegraphics{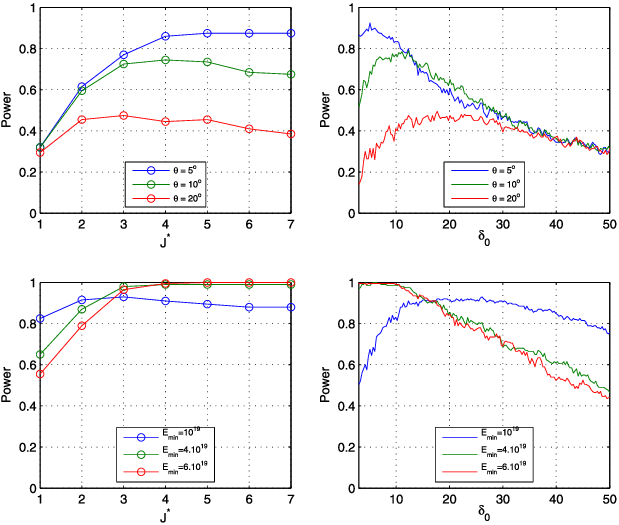}

\caption{The empirical power associated with the \multiple{} (left
column) and \twopc{} (right column) procedures with respect to their
key parameters $J^*$ and $\delta_0$, respectively. The prescribed
levels of the tests are 5\%. The three models under consideration in
the first row are provided by the alternative $\hypalt{a}$ with
$\theta
= 5^\circ, 10^\circ$ and $20^\circ$.
On the second row, the three alternative models are $\hypalt{c}$ with
$n_s = 500$, and $E_{\min} = 10^{19}$, $4\times10^{19}$ or $6\times10^{19}$~\textup{eV}.
The number of observations is $n=100$ everywhere.}
\label{fig:robustnessH1andH3}
\end{figure}

On Figure \ref{fig:robustnessH1andH3} we have plotted the estimated
power of the tests against different alternatives $\hypalt{a}$ or
$\hypalt
{c}$, and for different parameters for the methods. In the case of
$\hypalt{a}$, the first line of the figure shows that the optimal
$\delta
_0$ is indeed related to the parameters $\theta$ of the alternative.
However, when dealing with alternatives such as $\hypalt{c}$ (second line
of Figure \ref{fig:robustnessH1andH3}) that give rise to structures at
different ``scales'', the optimal choice of $\delta_0$ is not clear.
By observing the large variations of the power of the \twopc{}
procedure with respect to $\delta_0$ in both cases, one can conclude
that this procedure should incorporate a data-driven selection of
$\delta_0$ to be truly efficient.

\begin{figure}

\includegraphics{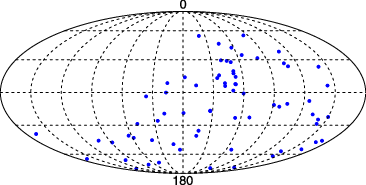}

\caption{The 69 arrival directions of cosmic rays with energy above 55 \textup{EeV}
and detected by the Pierre Auger Observatory up to 31 December 2009
[\citet{2010APh....34..314T}]. Their distribution is obviously nonuniform,
due to the incomplete coverage function of the instrument that is
described in Figure~\protect\ref{fig:auger_exposure}. Anisotropy tests actually
compare the empirical distribution to the exposure function (see text
for details).}
\label{fig:auger69}
\end{figure}

The situation is strikingly different for the needlet methods. One can
observe from the left column of Figure \ref{fig:robustnessH1andH3} that
the power reaches some \emph{plateau} after ${J^{\star
}}>J_{\min}$ in a
very consistent way across the different alternatives. This robustness
is a strong point of those methods. The dependence in $n$ is quite weak
too. For instance, taking ${J^{\star}}= 4$ leads to a
small loss of
efficiency uniformly with respect to the best choice for each given
situation of sample size and model.

\def\plugin{\multirow{3}{*}{\plugintxt}}
\def\multiple{\multirow{3}{*}{\textsc{Multiple}}}

\def\plugin{\textsc{PlugIn}}
\def\multiple{\textsc{Multiple}}

\section{Analysis of Auger data}
\label{sec:analysis-auger-data}

We have run the previous tests on the Auger public data made available
by the \citet{2010APh....34..314T}. It is composed of 69 arrival
directions of cosmic rays with energy above 55 EeV and detected by the
Pierre Auger Observatory between 1 January 2004 and 31 December 2009.
Those directional events are plotted on Figure \ref{fig:auger69}. The
distributions of the tests under study for $n=69$ and under the null
hypothesis have been evaluated by Monte Carlo simulation of length 10.000.

Along with the detection of a correlation between cosmic rays'
directions and catalogues of potential sources, the Pierre Auger
collaboration already performed a catalogue-free test for anisotropy
with no reference to any catalogue, using the \twopc{} procedure. As
noticed earlier, the critical value for this method is the choice of
$\delta_0$ in (\ref{eq:3}). The $p$-value of this test for the 69
UHECRs data set reaches a minimal value of
\[
\mbox{$p$-value}({\twopc{}}) \simeq0.008
\]
around $\delta_0 \simeq10.7^{\circ}$. Recall that in order to be
interpretable as a classical $p$-value for a single hypothesis testing,
this $p$-value should be computed from an out-of-the-sample
prescription of $\delta_0$, which is not the case here. Then this
$p$-value strongly exaggerates the significance of the detection.
Indeed, as already noticed in [\citet{2010APh....34..314T}], we computed
that the fraction of isotropic simulations that are as nonisotropic as
the real data at \emph{some} angle between $4^\circ$ and $14^\circ$ is
as high as 10\%. We have also computed that
\[
\mbox{$p$-value}({\efron{}}) \simeq0.07.
\]
The $p$-values of Table \ref{tab:pvalaugerneed} are the $p$-values
computed from the Pierre Auger data set for our \multiple{} and
\plugin
{} procedure.

\begin{table}
\caption{$P$-values of the \multiple{} and the \textsc{PlugIn} tests for
Auger data ($n = 69$)}\label{tab:pvalaugerneed}
\begin{tabular*}{\textwidth}{@{\extracolsep{\fill}}lcccccc@{}}
\hline
&\multicolumn{3}{c}{\textsc{\textbf{Multiple}} \textbf{test}}&\multicolumn{3}{c@{}}{\textsc{\textbf{PlugIn}}
\textbf{test}}\\[-6pt]
&\multicolumn{3}{c}{\hrulefill}&\multicolumn{3}{c@{}}{\hrulefill}\\
$\bolds{\bandmax}$ & \multicolumn{1}{c}{$\bolds{p=1}$} & \multicolumn{1}{c}{$\bolds{p=2^*}$} & \multicolumn{1}{c}{$\bolds{p=\infty}$}& \multicolumn{1}{c}{$\bolds{p=1}$} &
\multicolumn{1}{c}{$\bolds{p=2^*}$} & \multicolumn{1}{c@{}}{$\bolds{p=\infty}$} \\
\hline
1 & 0.957 & 0.788 & 0.387 & 0.956 & 0.958 & 0.397\\
2 & 0.051 & 0.112 & 0.035 & 0.033 & 0.037 & 0.036\\
3 & 0.118 & 0.050 & 0.004 & 0.017 & 0.008 & 0.005\\
4 & 0.434 & 0.046 & 0.003 & 0.017 & 0.008 & 0.008\\
5 & 0.227 & 0.095 & 0.624 & 0.017 & 0.008 & 0.008\\
6 & 0.762 & 0.045 & 0.341 & 0.017 & 0.008 & 0.008\\
\hline
\end{tabular*}
\end{table}

For the \multiple{} test, the $p$-value is defined as the proportion of
draws (under the null) that have a higher single test statistic in at
least one value of $j \in\{1,\ldots,{J^{\star}}\}$. The
resulting
$p$-value is quite sensitive to the choice of the highest band
$J^{\star}$, except if one uses the $\Lset^2$-norm. %
Note that if we take the $\Lset^2$ norm and the theoretical
${J^{\star}}=
2$ given by the expression (\ref{eq:expressionbandmax}), the results
for the \multiple{} test are not statistically significant. But the
Monte Carlo simulations suggest that this theoretical choice of
${J^{\star}}$ is not optimal for small to medium sample
size, being too small.

The \plugin{} is more stable and consistently considers that the Auger
data is
significantly nonisotropic. The almost constant $p$-values in this case
are the
consequence of a hard thresholding rule in (\ref{eq:defhatf})\vspace*{1pt} that
cancels all
the estimated coefficients $\hat\beta_{j,k}$ as soon as $j \geq3$
for this
data set. This may in turn give a rule-of-thumb rule to define a data-driven
$J^{\star}$ for the multiple test.

To conclude on this important data set and this methodology, it appears
that the needlet methods find a stronger statistical evidence of some
kind of anisotropy in the Pierre Auger data. More realistic
alternatives and more simulations can help to choose the ${J^{\star}}$
parameter of the \multiple{} procedure and additional parameters of the
\plugin{} approach.

\section{Conclusion}

In this paper we have investigated the problem of the detection of
anisotropy of directional data on the unit sphere, with an application
to the analysis of ultrahigh energy cosmic ray events as observed with
a detector such as the Pierre Auger Observatory. It was important to
consider samples whose sizes are comparable to the sizes of the data
sets that are available nowadays for cosmic rays scientists (about 25
at the beginning of this work, about a hundred now). Although we are
mainly interested in small sample performances, we have proposed a
multiple test approach based on a multiresolution analysis of the data,
which could hopefully be proved to be asymptotically optimal in the
minimax sense, a~well-known pessimistic framework.

We have proposed, and tested on various simulated data sets, two
methods using the decomposition of the directional data onto a frame of
spherical needlets. Their performance has been compared to other (more
specific) approaches based on the nearest neighbour and on the
two-point correlation function. The simulation shows that the
needlet-based methods perform comparatively very well in various
situations. They are competitive with the existing method at a small
sample size, and tend to outperform them from a moderate sample size.
Moreover, the ``omnibus'' property of the needlets method is
interesting for the problem at hand, in which the type of possible
anisotropy (the class of alternative) is not really well known a
priori. In addition, a multiple test based on the use of spherical
needlets offers a good opportunity to extend the method of detection of
anisotropies with not only multiplicity in the scales tested, but also
in ranges of energy of the incoming particles. Indeed, while in this
work we have used the energy level as a simple threshold, one could
instead implement a detection using the joint directional-energy
information---allowing thus to simultaneously extract information from
the highest energy cosmic rays, which are not deflected much by
Galactic and extragalactic magnetic fields, and also from lower energy
events, more deflected but much more numerous. In light of our
simulations on an energy level-dependent model, the multiscale approach
could lead to stronger conclusion using the CR data that are not yet
made public by the Pierre Auger Observatory.

As in any nonparametric method, there is at least one parameter to be
tuned, often by hand or using more sophisticated data-driven methods
such as cross-validation. In the needlet methods one can tune several
parameters (shape of the needlets, highest scale $J^{\star}$--- although
there is an asymptotic formula for it, thresholds on the coefficients
in the \plugin{} approaches, thresholds on the individual tests in the
\multiple{} procedure, power norm). It is plausible, however, that a
large range of possible choices for most of these parameters give
comparable performance.

Although we have used needlets that are compactly supported windows in
the harmonic space, it may be arguable that they are not the most
appropriate tool. One could consider, as an alternative, better
spatially concentrated functions [see, e.g., \citet{lan:marinucci:2009}, such as the Mexican needlets]
or, in general, try to optimize the
needlet window function given prior knowledge of the physical problem
and of the expected properties of anisotropic distributions of the
cosmic ray direction of incidence. In this spirit, it would be
interesting to consider directional wavelet such as curvelets or
ridgelets [see \citet{starck+2006}] to test for specific strip-like
alternative densities. It is also possible to consider nondyadic
needlets. The choice of $\cB\in(1,2)$ allows a finer coverage of the
frequency line. The numerical results presented here have not taken
this benefit into full account, and whether significantly higher power
can be obtained by optimizing this number remains to be investigated.

Finally, in addition to the aforementioned possible extensions of our
methods, we want to stress that the work presented here also opens the
way to two lines of future investigations, one on the applications side
and one more theoretical. On the experimental side, it will be of much
interest to apply the method on larger data sets (for instance, by
lowering the energy threshold to increase the available sample size).
On the theoretical side, the validation of the approach has to be
investigated on the basis of some theory in the minimax framework it is
designed for.

\section*{Acknowledgments}
For the numerical part of this work, we acknowledge the use of the
\textsc{Healp}ix package [\citet{gorski+2005}] and of the SphereLab (an
\textsc{Octave} interface to \textsc{Healp}ix package, needlet
transforms and utilities). We thank Dmitri Semikoz for useful
discussions concerning high energy cosmic rays, and Maude le Jeune,
Jean-Fran\c{c}ois Cardoso and Fr\'{e}d\'{e}ric Guilloux for making
available some of the tools used for the computational aspects of the
present work. We thank the anonymous referees and Associate Editor for
their suggestions and remarks that significantly improved the
presentation of this paper.

\begin{supplement}[id=suppA]
\stitle{Supplement to ``Testing
the isotropy of high energy cosmic
rays with
spherical needlets''}
\slink[doi]{10.1214/12-AOAS619SUPP} 
\sdatatype{.pdf}
\sfilename{aoas619\_supp.pdf}
\sdescription{In the supplement, we recall
 the construction of the needlet decomposition on
the sphere, and discuss
its practical usage. We also complete the Section \ref{sec:monte-carlo-exper} of this paper with more
results obtained from Monte-Carlo simulations.}
\end{supplement}

%
%

%

%

\printaddresses

\end{document}